\documentclass[%
 reprint,
 amsmath,amssymb,
 aps,
]{revtex4-2}

\usepackage{graphicx}% Include figure files
\usepackage{graphics}% Include figure files
\usepackage{dcolumn}% Align table columns on decimal point
\usepackage{bm}% bold math
\usepackage{mhchem}
\usepackage{siunitx}
\usepackage{xcolor}

\newcommand{\sln}{SrLn$_{2}$O$_4$}
\newcommand{\sho}{SrHo$_{2}$O$_4$}

\newcommand{\stm}{SrTm$_{2}$O$_4$}

\newcommand{\tm}{Tm$^{3+}$}

\begin{document}
\preprint{}

\title{On the pseudo-doublet ground state of the non-Kramers compound \stm \, and its frustrated antiferromagnetic interactions}
\author{D. L. Quintero Castro$^{1}$}
\email[]{diana.quintero-castro@psi.no}
\thanks{current address: Laboratory for Neutron Scattering and Imaging, Paul Scherrer Institute, 5232 Villigen PSI, Switzerland.}
\author{A. Bhat Kademane$^{1}$}
\author{M. Pregelj$^2$}
\author{R. Toft-Petersen$^{3,4}$}
\author{D. G. Mazzone$^{5}$}
\author{G. S. Tucker$^{6}$}
\author{C. Salazar Mejia $^{7}$ }
\author{J. Gronemann $^{7}$ }
\author{H.-F. Li$^{8}$}

\affiliation{$^{1}$ University of Stavanger, 4036 Stavanger, Norway\\
$^2$ Jožef Stefan Institute, Jamova c. 39, 1000 Ljubljana, Slovenia \\
$^{3}$ Department of Physics, Technical University of Denmark, Fysikvej, 2800 Kongens Lyngby, Denmark \\
$^{4}$ European Spallation Source ERIC, P.O. Box 176, SE-221 00, Lund, Sweden\\
$^{5}$ PSI Center for Neutron and Muon Sciences, 5232 Villigen PSI, Switzerland\\
$^{6}$ European Spallation Source, Data Management and Software Centre, DMSC Asmussens Alle 305, DK-2800 Kongens Lyngby, Denmark\\
$^{7}$ Dresden High Magnetic Field Laboratory (HLD-EMFL), Helmholtz-Zentrum Dresden-Rossendorf (HZDR), Dresden, 01328, Germany  \\
$^8$Institute of Applied Physics and Materials Engineering, University of Macau, Avenida da Universidade, Taipa, Macao SAR 999078, China  
}

\date{\today}

\begin{abstract}

Here we present experimental evidence of the pseudo-doublet ground state of the non-Kramers compound \stm, based on specific heat, magnetic entropy and electron paramagnetic resonance. We demonstrate that the two crystallographic \tm \, sites give rise to distinct single-ion anisotropies, and by extension, \stm \, hosts two magnetic sublattices.
Inelastic neutron scattering reveals low-lying dispersing crystal-field excitations, which we modelled using an effective charge model and mean field random phase approximation. The extracted magnetic exchange interactions are both antiferromagnetic and frustrated for both sublattices. Interchain magnetic exchange interactions are negligible. The strength of the magnetic exchange interactions in relation to the size of crystal field energy gaps, together with the frustration and low dimensionality, force the system to remain paramagnetic down to the lowest experimentally reachable temperature despite the pseudo-doublet nature of its ground state.  

\end{abstract}

\pacs{75.25.-j, 75.40.−s, 75.50.−y, 75.30.Gw, 75.40.-s}    

\maketitle

\section{Introduction}

Non-Kramers ions usually have a non-magnetic singlet ground state, however, when enforced by specific site symmetries, they can have a pseudo-doublet ground state, by the admixing of dipolar and electric quadrupolar moments, presenting an effective spin-1/2 (S$=1/2$) with strictly Ising anisotropy and intrinsic transverse field \cite{Rau,TmMgGaO4_theory}. Experimental realizations of this ground state have been reported for KTmSe$_2$ \cite{KTmSe2}, TmMgGaO$_4$ \cite{TmMgGaO4_2020} and NaTmTe$_2$ \cite{NaTmTe2}. In these compounds, the \tm \, ions are arranged in geometrically frustrated patterns in quasi-two-dimensional structures. They all show dispersing crystal field (CF) excitations at low energies with various gap sizes. From these, the only compound developing long-range magnetic order is TmMgGaO$_4$, also showing the smallest gap size. 

Here, we present the case of a non-kramers-based compound \stm. It belongs to the family of compounds \sln \, (Ln: Yb, Er, Ho, Tb, Dy and Tm). Their atomic structure comprises two non-equivalent crystallographic sites for the lanthanide ion both with C$_s$ point symmetry, forming two zigzag chains running along the $c$-axis and making a distorted honeycomb lattice on the $ab$-plane as shown in Fig. \ref{fig:Struct}. All the atoms are located on the Wyckoff site 4c of the space group
Pnam. These compounds show either low-, three-dimensional or coexistent types of magnetic order in the mK temperature range (for a review see Ref.\cite{Petrenko2014}). A second non-Kramers-based ion compound in the family, \sho, shows two coexistent short-range magnetic orders \cite{young2013} and is described as a S$=1$ system with Ising anisotropy \cite{srho2o4}.

\stm \, does not develop any magnetic order down to $65$\,mK \cite{Haifeng2015} and like the Tm compounds mentioned above, shows low-lying dispersing CF excitations. In this work, we investigate these excitations by using inelastic neutron scattering (INS) results, which we modelled by combining an effective charge crystal-field model and mean-field random phase approximation. We extracted the magnetic exchange interactions, which indicate the compound is made out of two highly frustrated independent low-dimensional magnetic sublattices. The model presented here is based on our published crystal field parameters obtained using an effective charge model (see Ref. \cite{kademane2021}) which revealed different single ion anisotropy for each crystallographic site.  In addition to INS, we present specific heat, magnetic entropy and electron paramagnetic resonance results. These results agree with the description of mixed anisotropies and show that the actual effective ground state is a pseudo-doublet for both sites as in the case of KTmSe$_2$, TmMgGaO$_4$ and NaTmTe$_2$.

\section{Experimental Details}

The neutron multiplexing spectrometer CAMEA at PSI \cite{CAMEA} was used to measure the low energy excitation spectra ($<4$\,meV) in \stm. Data reduction and analysis were performed using the software MJOLNIR \cite{MJOLNIR}. For this experiment, a large cylindrical single crystal of mass $\sim$ 3.2\,g was mounted on an aluminium sample holder in two orientations to access both the dynamics in the (0KL) and the (H0L) plane. All measurements were performed at $1.6$\,K. 
Continuous wave electron paramagnetic resonance (EPR) measurements were performed in X-band (frequency of about 9.4 GHz) on a Bruker E500 spectrometer equipped with a Bruker ER 049X microwave bridge, and a Super-high-Q resonator ER 4122 SHQ. The single-crystal sample with a mass of $12.4$\,mg was mounted on a quartz sample holder. An Oxford Cryogenics continuous-flow liquid He cryogenic system ensured temperature stability better than $0.1$\,K. 
Specific heat measurements below $1$\,K were performed using a homemade setup installed in a dilution refrigerator, following the heat pulse method with $1-3\%$ heating pulses at HZDR on the same single crystal sample used for the EPR measurements.

All measurements were made in the single crystal used for the diffraction analysis shown in Ref.\cite{Haifeng2015}.

\section{Results}
\subsection{Mean field description of dispersing crystal field levels (T$>$J$_{ij}$)}

Multiple neutron incident energies were used to obtain the full INS spectra with $0.5$ to $4$\,meV energy transferred. Figs. \ref{fig:camea}(a, c and e) show the key results in high symmetric directions. The spectra consist of CF-dispersing modes divided into two gapped bands  expanding from $\thicksim 0.77 - 1.82$\,meV and $\thicksim 3 - 3.98$\,meV.
The dispersion is significant along $\langle 00L\rangle$ and is reminiscent of the one proposed by Müller and Mikeska for uncoupled $S = 1/2$ antiferromagnetic zigzag chains with magnetic exchange interactions for the chain rung one and a half times stronger than those along the chain legs \cite{Muller}. 
The two bands show similar characteristics with minima at $L\approx 0.36, 0.66$ (white arrows in Fig. \ref{fig:camea}(c)) and maxima at the gamma point.    The dispersion is negligible along $\langle H00\rangle$  and $\langle0K0\rangle$.  
 All modes are resolution limited; no broadening, nor diffuse scattering has been detected. 

\begin{figure}[htb!]
\centering
        \includegraphics[width=0.5\textwidth]{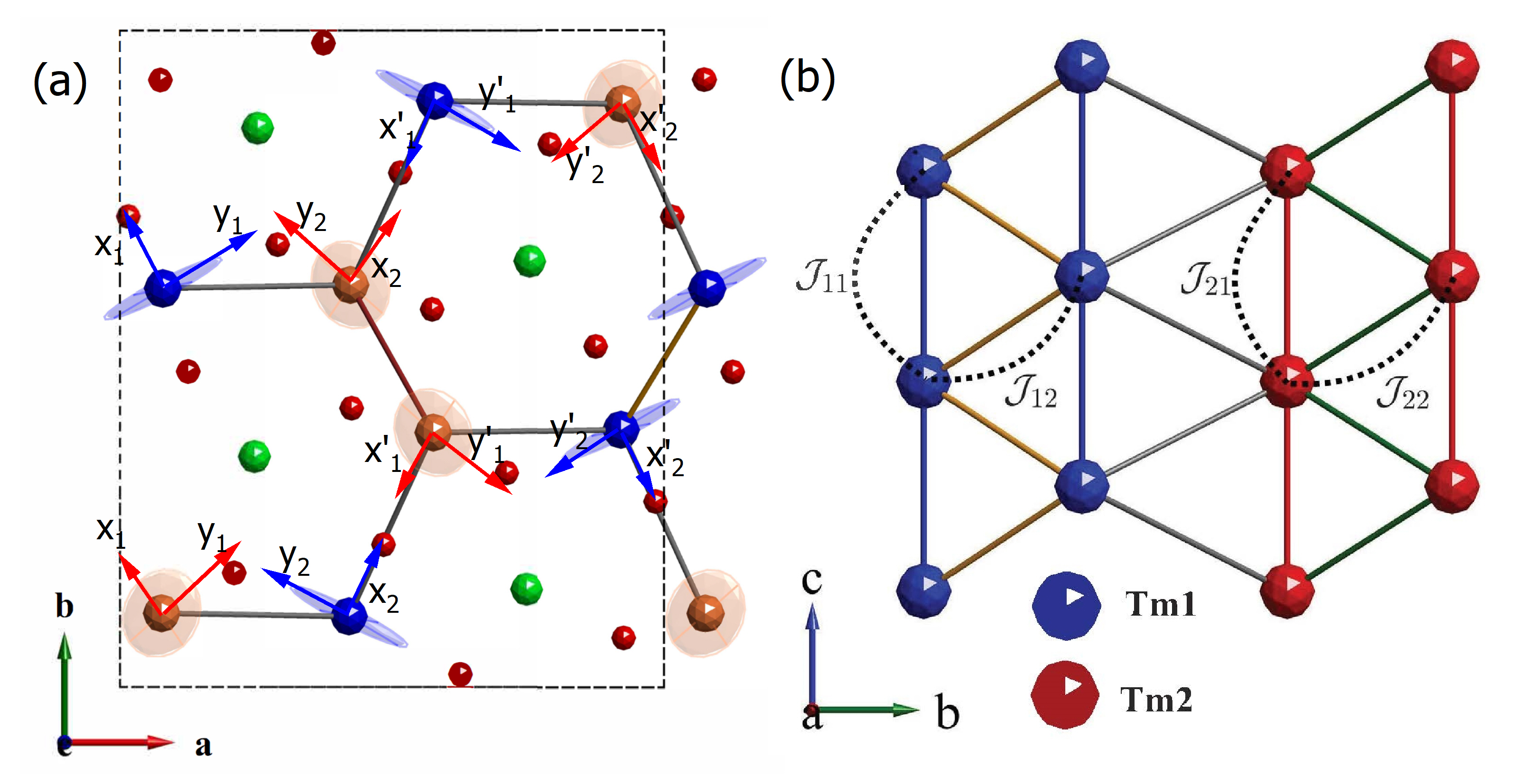}
        \caption{Graphic representation of the atomic structure of \stm \, in the $ab$-plane (a) and $bc$-plane (b). Tm1 sites are shown in blue and Tm2 in red/orange. Strontium ions are shown in green and oxygen in red (a) shows the single ion anisotropy estimated from the CF calculations along with axes (blue-Tm1, red-Tm2) that define the symmetry of each site. (b) shows the zig-zag chains along with the magnetic interactions considered in Eq.\ref{Eq:RPA}. Drawings made using SpinW \cite{SpinW}.}
    \label{fig:Struct}
\end{figure}

Because the ground state is not ordered, we modelled the dispersive CF modes using random phase approximation (RPA) theory implemented within the \emph{mcdisp} module in \texttt{McPhase} \cite{Mcphase, Rotter2017,Jensen1991} by assuming the hypothesis that our CF model in Ref. \cite{kademane2021} assigns well the origin of each of these bands to two independent zigzag chains, one for Tm1 and one for Tm2 (as shown in Fig. \ref{fig:Struct}).  In this model, both \tm \, ions have a singlet ground state and a first excited level with a dominating total angular momentum, $\mathbf{J}=1$, however, with a fair level of up-mixing. From this CF model, the gap to the first excited level is $\Delta_1= 3.1$\,meV for the first site, Tm1, and $\Delta_2=0.9$\,meV for the second site, Tm2. These gaps values agree to a large extend with the CAMEA data shown in Fig. \ref{fig:camea}. This model supports the assignment of the energies of the first excited state to each of the sites with those using site-selective laser spectroscopy published by another group as found in Ref. \cite{NIKITIN2024120616}, reporting a gap from the ground singlet state to the first excited state of $3.69$\,meV for Tm1 and $1.12$\,meV. The slight difference between these gaps values can arise from differences in energy resolution between the different technique and sample differences between powders vs single crystal as well as the full occupation of Tm ions in the \sln \, atomic structure.

According to our CF model Tm1 has an Ising type anisotropy whereas Tm2 can be considered as having a planar single-ion anisotropy. Both the anisotropy axis/plane lay on the crystallographic $ab$-plane as shown in Fig. \ref{fig:Struct}(a). Individual axes for each Tm site show that there are four different magnetic sites for each crystallographic site, sets labeled 1 and 2 related by a two-fold rotoinversion and prime and not prime subsets related by mirror symmetry. Therefore there are various orientations of the long anisotropy axis.
%In the case of Tm1, all magnetic sites are related by a rotation around the c-axis. As for the Tm2 site, the four sites are divided into two sets, within a set, two sites are related by a 4-fold roto-inversions around the c-axis, and a 2-fold rotation between both sets.
The single-ion anisotropy considerations can be included in the RPA model using the exchange tensor as described in Ref.\cite{Rotter2000}. Thus, the following Hamiltonian was used to model the system.

\begin{equation}
\begin{aligned}
\mathcal{H}_n=\sum_{Tm1 , l m} B_{l}^{m} O_{l m}\left(\mathbf{J}^{n}\right)
-\frac{1}{2} \sum_{Tm1: n n^{\prime}} \mathcal{J}_{n n^{\prime}} \mathbf{J}_z^{n} \mathbf{J}_z^{n^{\prime}} \\
+ \sum_{Tm2, l m} B_{l}^{m} O_{l m}\left(\mathbf{J}^{n}\right)
-\frac{1}{2} \sum_{Tm2:n n^{\prime}} \mathcal{J}_{n n^{\prime}} \mathbf{J}_{xy}^{n} \mathbf{J}_{xy}^{n^{\prime}} .
\label{Eq:RPA}
\end{aligned}
\end{equation}

The Hamiltonian includes two crystal field terms for both Tm$i$ sites, where $O_{l m}$ is the Stevens operator and the $B_{l}^{m}$ the crystal field parameters and two terms corresponding to the magnetic exchange interactions for both ions, with $\mathcal{J}_{n n^{\prime}}$ being the exchange tensor. Here, we ignore two-ion anisotropy, typically presented as `Racah operators' \cite{Jensen1991}, multi-pole interactions and anti-symmetric exchange. Most importantly, notice that these are essentially two Hamiltonians for two non-interacting magnetic sublattices. 

\begin{figure}[htb!]
  \centering
\includegraphics[width=0.51\textwidth]{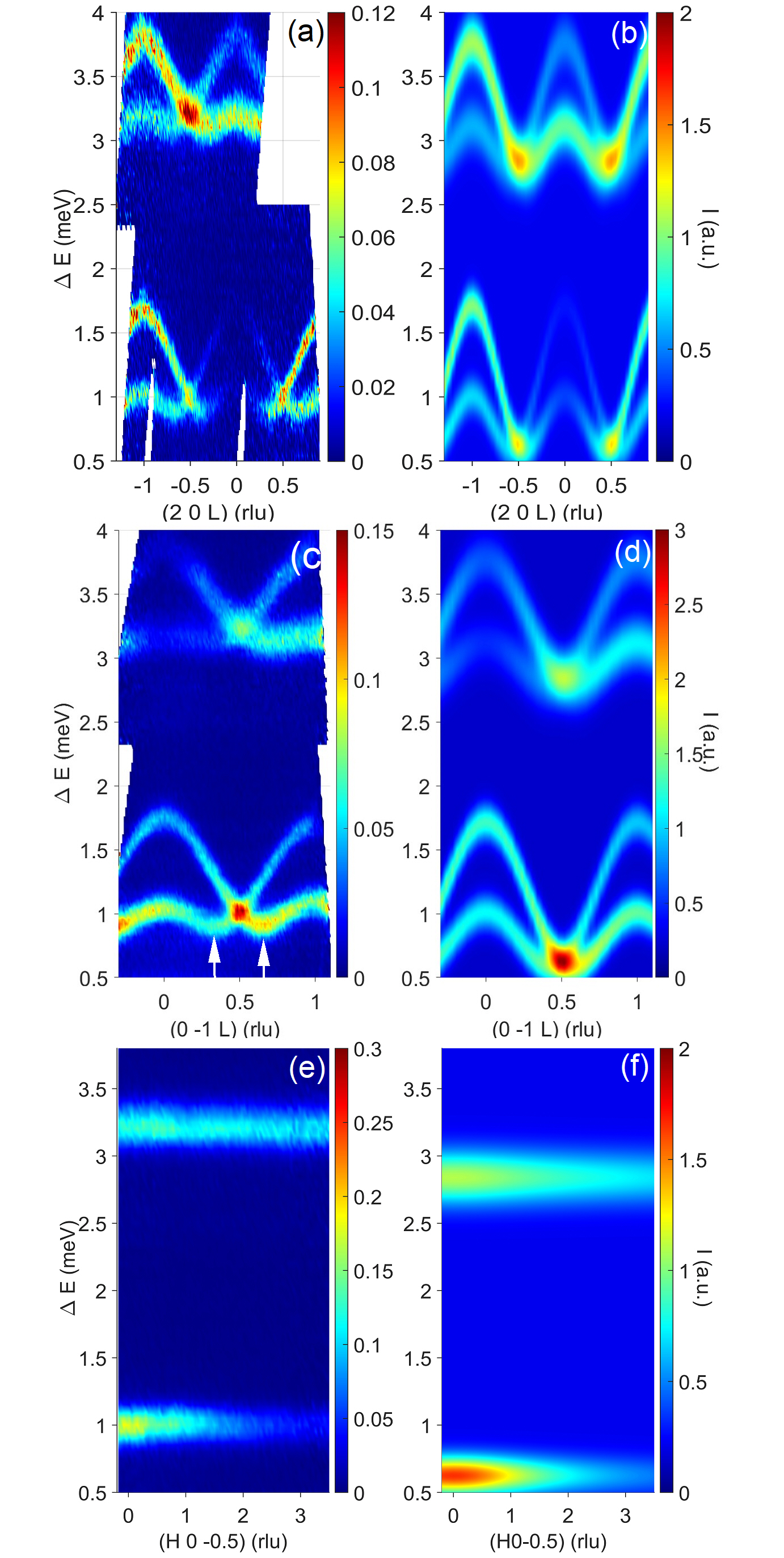}
%\smallskip
%\includegraphics[width=0.51\textwidth]{01L.jpg}
%\smallskip
%\includegraphics[width=0.51\textwidth]{H0m0p5T.jpg}
\caption{ Comparison between INS (a, c, e) and RPA (b,d,f) spectra for three different directions in reciprocal space at $1.6$\,K. Colours indicate intensity in different arb. units for data and simulation. White arrows in (c) indicate the spectra minima.}
\label{fig:camea}
\end{figure}

Figure \ref{fig:Struct}(b) shows and identifies the assumed exchange paths for both Tm chains included in Eq. \ref{Eq:RPA}, only two interactions per chain are considered -rung and leg-.   The exchange interactions were fitted to the experimental data using a simulated annealing procedure called \emph{simannfit} in \texttt{McPhase} \cite{Mcphase, Rotter2017}. Table \ref{tab:Itn} presents the exchange interactions obtained from the fitting and the corresponding distances. The simulated spectra are shown in figures \ref{fig:camea}(b), \ref{fig:camea}(d) and \ref{fig:camea}(f).  The simulation agrees very well with the observed dispersion relation and relative dynamical structure factors, while the overall bandwidth (gap) is overestimated (underestimated) by $\thicksim0.4$\,meV.

Other Hamiltonian models were considered, for instance, Dzyaloshinskii-Moriya interactions for Tm2 along the rung of the zig-zag chain were included but showed no effect on the calculated S(Q,w). Including interchain interactions in the Hamiltonian above generates $12$ modes below $4$\,meV, due to the spatial differences in exchange paths.

The obtained two-ion magnetic exchange interactions are all antiferromagnetic. Fig. \ref{fig:Struct} and Table \ref{tab:Itn} show exchange paths, distances and strengths. Nearest and next-nearest neighbour distances are very similar, as well as the strength of these interactions. The ratios between these interactions are for Tm1 $\alpha = \mathcal J_{12}$/$\mathcal J_{11} \approx$ $1.06$, while Tm2 has a ratio of $\mathcal J_{22}$/$\mathcal J_{21} \approx$ $1$. 
The strength of the magnetic exchange interactions is much smaller than the crystal field gaps. However, the interaction strengths can be scaled up by using a different value for the total orbital angular momentum or by using an effective spin value instead.

%Following the Mean Field RPA theory as indicated in Eq. 7.1.6 in Ref. \cite{Jensen1991}, a critical parameter can be calculated as $R_0=2M^2\mathbf{J}/\Delta$, where $M$ is the numerical value of the matrix element of $\mathbf{J}$ between the two states. The calculated $R_0$ value for \stm \, is $0.2$, far from 1, implying that it is impossible to drive this system through a second-order thermal phase transition to an ordered state without a magnetic field. 

\begin{table}[!htbp]
\setlength{\tabcolsep}{11pt}
\renewcommand{\arraystretch}{1.5}
\caption{RPA resulting magnetic exchange interactions with their respective distances and gaps from the CF model.} % title name of the table
\centering % centering table
\begin{tabular}{c c  c  c c } % creating 10 columns
\hline\hline % inserting double-line
 Site&\multicolumn{2}{c}{$\mathcal J$(meV)} &$r_i$(\AA) & 
\\ \hline%[0.5ex]
%\hline% inserts single-line
% Entering 1st row
Tm$1$&  $\mathcal J_{11}$    & 0.016(3)   &   3.3811 & $\Delta_1$= 3.1\,meV\\ 
    &  $\mathcal J_{12}$    & 0.017(4)    & 3.4212 & $\alpha_1\approx$1.06 \\ \hline
Tm$2$&  $\mathcal J_{21}$    & 0.007(1)   &  3.3811 & $\Delta_2$= 0.9\,meV \\ 
     &  $\mathcal J_{22}$    & 0.007(1)    &  3.4860 & $\alpha_2\approx$1  \\ \hline
\hline % inserts single-line
\end{tabular}
\label{tab:Itn}
\end{table}

\subsection{Pseudo-doublet ground state, T$\approx \mathcal{J}_{ij}$}

\begin{figure}
      \centering
	  \includegraphics[width=0.5\textwidth]{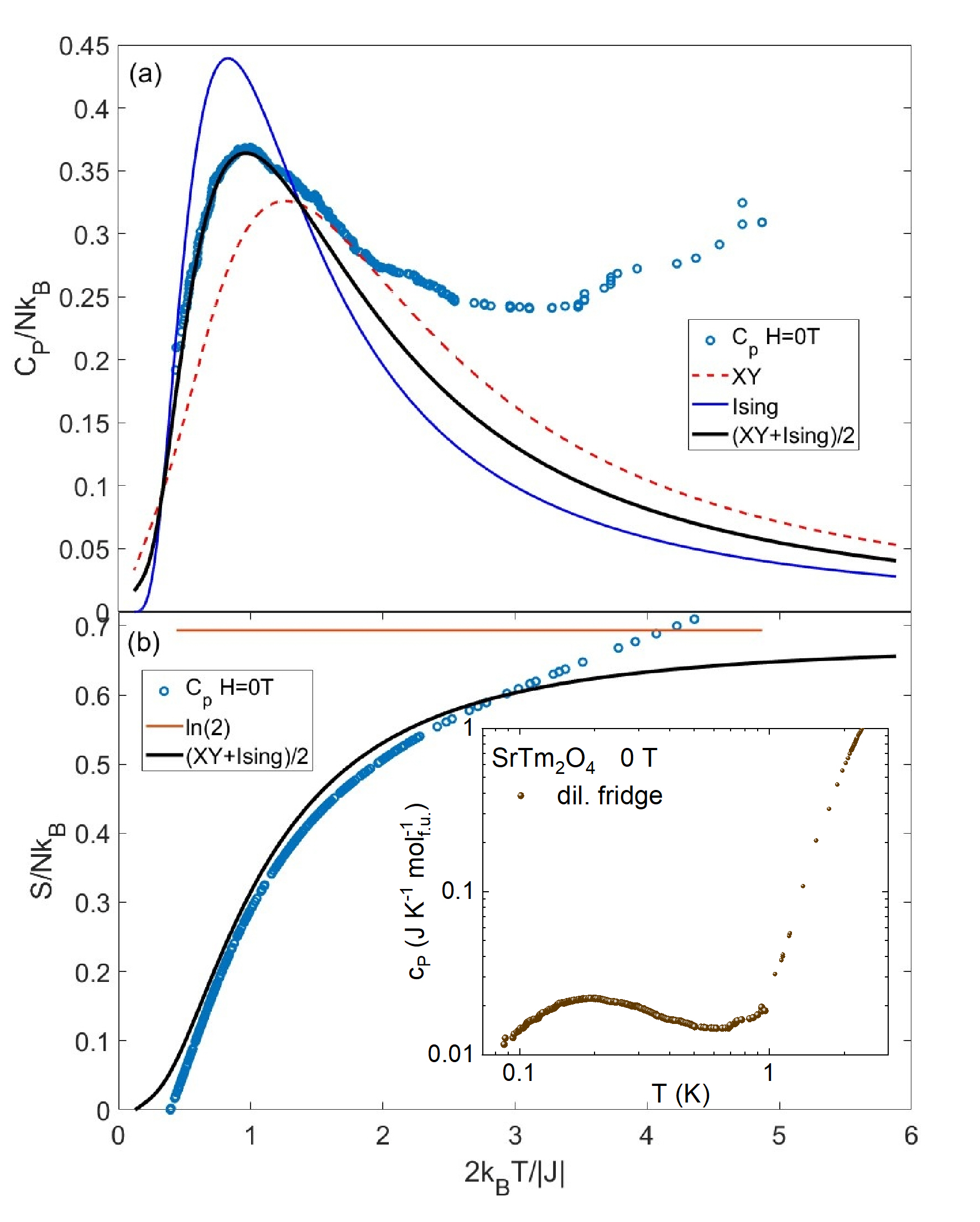}
         \caption{(a) Measured and calculated reduced specific heat per ion as a function of reduced temperature, shown here for the XY, Ising and XY-Ising average. (b) Derived magnetic entropy from the specific heat -measured and calculated- shown in (a). $ln(2)$ is shown as a reference. Insert in (b) shows the specific heat results as measured.}
  	\label{fig:Cp_S}
\end{figure}

In light of the RPA results, we have investigated the temperature dependency of the specific heat of \stm \, between $0.07$\,K and $1$\,K, the temperature range where these interactions are relevant ($0.19$\,K for Tm1 and $0.08$\,K for Tm2).  These results are shown in Fig. \ref{fig:Cp_S} as a function of reduced temperature. The data has been smoothed by taking the average of multiple measured points, resulting error bars corresponding to a $5\%$ of the specific heat value are not shown here. In this temperature range, the specific heat shows a broad maximum centred at $0.19$\,K.

The RPA results are consistent with \stm \, consisting of two non-interacting magnetic sublattices, one Ising and one planar. A 2D Ising system should show a lambda-type anomaly in the specific heat described by the Onsager integral \cite{Onsager}. The 1D cases for both XY and Ising exhibit broader features as calculated by Katsura in Ref \cite{Katsura} with the XY case showing a peak at slightly higher reduced temperatures than the Ising case. In Fig. \ref{fig:Cp_S}, we show along with the measured data, the calculated specific heat for both the 1D Ising and the 1D XY case following Ref \cite{Katsura}, as well as their average value. Here, we plot all as a function of the reduced temperature assuming a value for $\mathcal{J}$ of $0.034$\,meV ($0.39$\,K) for both sublattices. This exchange value is larger than the RPA-obtained exchange interactions in Table \ref{tab:Itn}, however, we argued that these values depend on the chosen value of $\mathbf{J}$. The averaged model fits the specific heat data remarkably well, suggesting an overestimation of the total angular momentum for both sublattices in Eq.\ref{Eq:RPA}. The specific heat model differs at higher reduced temperatures, when approaching $1$\,K, as the measured specific heat increases due to the crystal field levels ($\Delta_1= 3.1$\,meV$\approx 36$\,K and $\Delta_2=0.9$\,meV$\approx 10.4$\,K)  and phonon contributions as it was shown and modelled in Fig.1d of Ref. \cite{kademane2021}.

The corresponding magnetic entropy has been calculated both for the averaged model and the experimental data, this comparison is shown in Fig. \ref{fig:Cp_S} along with a line that indicates the value of $ln(2)$ which is reached for temperatures above $0.79$\,K ($>\mathcal{J}_{ij}$), which indicates that the ground state for both sites is a pseudo-doublet.

\subsection{Electron Paramagnetic Resonance}

Non-Kramers ions in high symmetry sites and pure non-degenerate singlet ground states do not interact with a magnetic field and thus do not show any EPR signal. However, if the symmetry is low enough it allows the admixing of eigenstates showing a pseudo-doublet behaviour, with a dipolar moment and therefore can display EPR signal. This is possible however with a sufficiently low energy gap between the ground and the first excited state ($\Delta$) but a large energy gap to the second exited state \cite{EPR_nonKramers}. 
EPR signals for \tm \, ions have been detected for trigonal site symmetries C$_{3v}$ and triclinic C$_1$, all of them reporting large g$_\parallel$ and zero g$_\perp$ \cite{EPR_Tm, YU_EPR}. In the pseudo-spin Hamiltonian notation, g$_\perp$ must be zero as S$_x$ and S$_y$ do not correspond to an observable \cite{EPR_nonKramers, Rau, TmMgGaO4_theory}. Thus, EPR is an ideal tool to prove whether an effective $S=1/2$ is a valid description of \stm \, as it is the case for KTmSe$_2$ \cite{KTmSe2}, TmMgGaO$_4$ \cite{TmMgGaO4_2020} and NaTmTe$_2$ \cite{NaTmTe2}.  

\begin{figure*}[htb!]
  \centering
\includegraphics[width=1\textwidth]{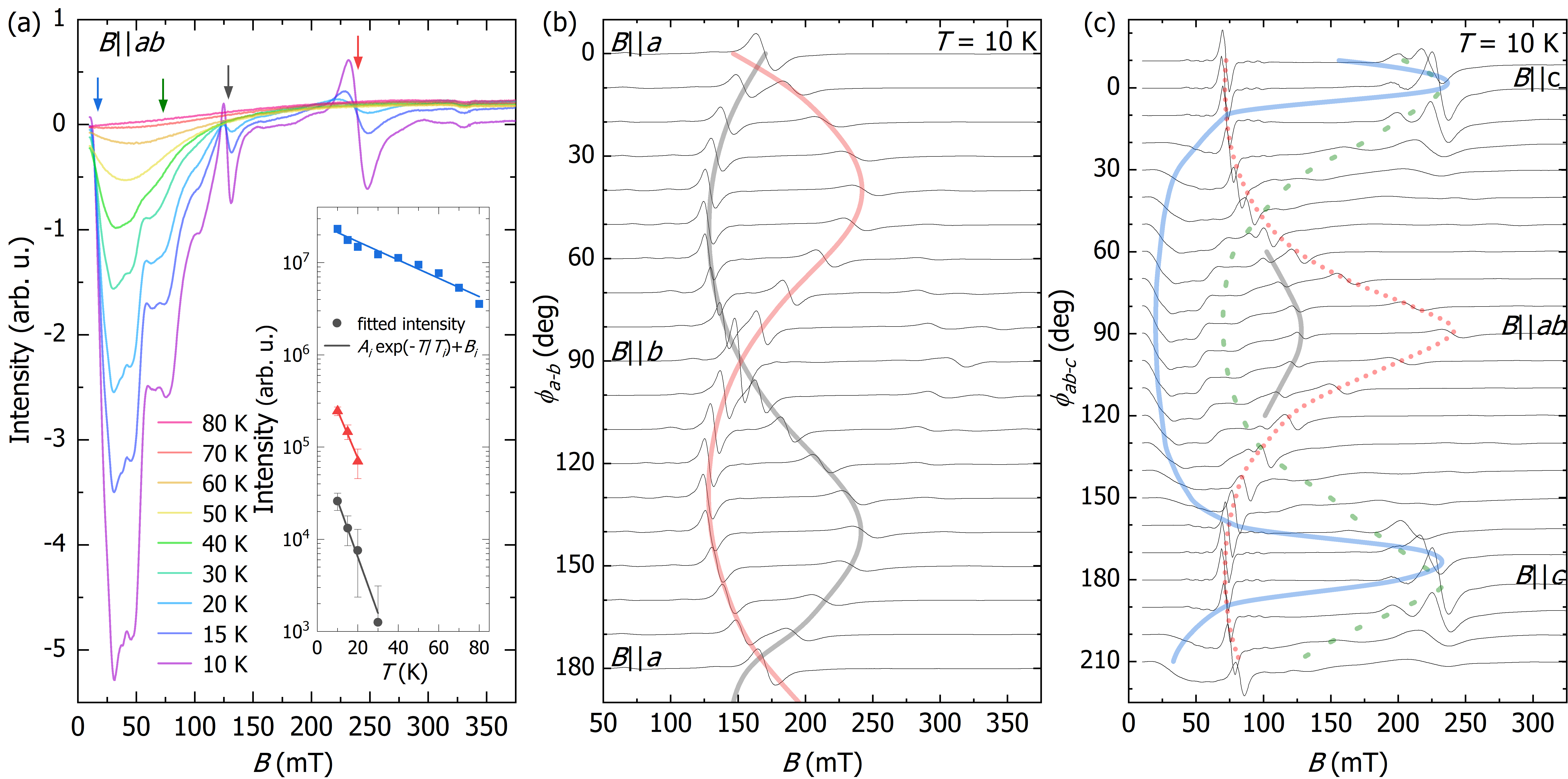}
\caption{(a) Temperature dependence of the EPR spectra for \stm \, for magnetic fields along the $[110]$ direction. The inset shows the derived intensities of the signals marked with arrows in the main figure. The solid line is a fit to an exponential function.  (b, c) Angular dependencies of the EPR spectra at $10$\,K with the magnetic field applied at various positions within (b) the $ab$- plane and (c) the $(HHL)$ plane. The lines are guides for the eyes, while their colors relate them with the signals marked by arrows in (a).}
\label{fig:EPR1}
\end{figure*}

The temperature dependencies of the EPR spectrum for magnetic fields applied along $[110]$ are shown in Fig. \ref{fig:EPR1}(a). Three main contributions are noticeable in these data sets, a low-temperature, low-field signal with a large g factor that develops below $100$\,K (blue arrow) and two contributions that develop at temperatures below $\sim 30$\,K with smaller g factors (black and red arrows). Much weaker contributions at higher fields are also detected but are not shown here. These are most likely related to transitions within higher CF levels where their intensity is associated with their population.  The derived temperature dependencies for the three key modes (marked with arrows with fits of the corresponding colour) are shown in the figure inset. The three signals decay exponentially with an exponent T$_1\approx 44(4)$\,K$\approx 3.7(3)$\,meV for the largest g component and T$_1\approx 8(1)$\,K$\approx 0.7(1)$\,meV and T$_1\approx 7.1(5)$\,K$\approx 0.61(5)$\,meV for the smaller g modes. If we assume these temperatures are linked to a gap to an excited state we can see how similar they are to $\Delta_1$ and $\Delta_2$, thus to extend our hypothesis, we assign the large g-factor signal to Tm1 (blue) and the two small g-factor signals to Tm2 (black and red).

\begin{figure*}[htb!]
  \centering
\includegraphics[width=1\textwidth]{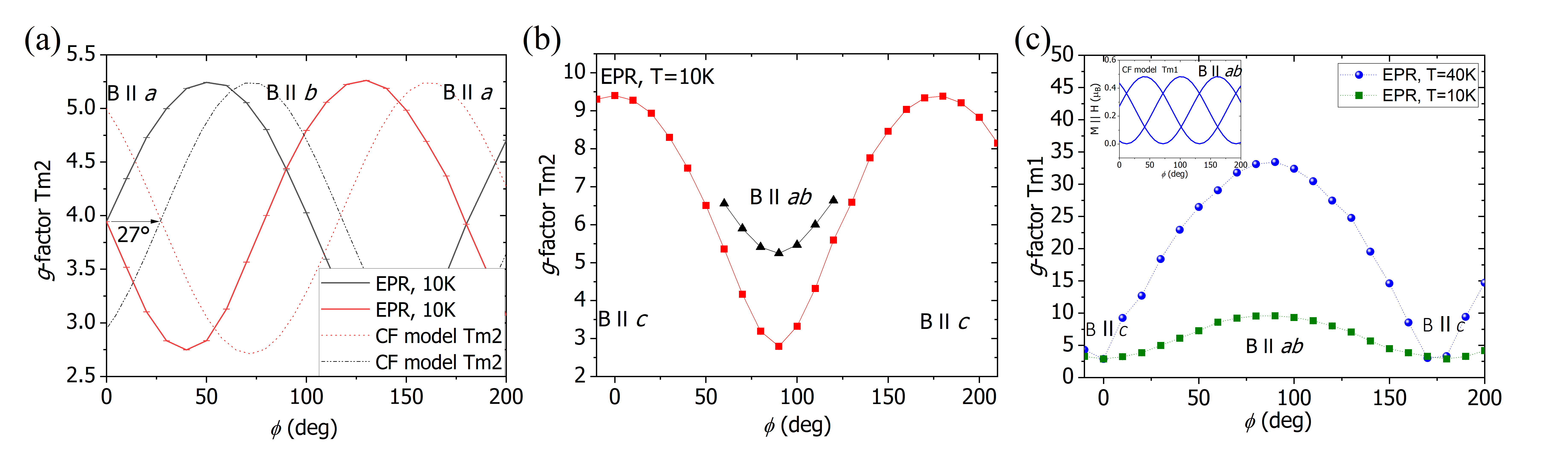}
\caption{Derived angular dependencies of the g factor values for Tm2 [(a) and (b)] and for Tm1 (c).  The calculated magnetic moment from our CF model is also plotted in (a) for the corresponding plane and shown as an inset in (c) for the perpendicular plane for each ion.}
\label{fig:EPR_fig2}
\end{figure*}

These differences in activation temperatures are actually experimentally convenient as we can separate the contributions of Tm1 and Tm2 by setting the experiment temperature. Consequently, we performed orientation-dependent measurements by rotating the sample within the $ab$-plane and the $(HHL)$-plane, as shown in Figs.\ref{fig:EPR1}(b,c). The angular dependencies of the extracted g-factors of the modes marked with black and red arrows in Fig.\ref{fig:EPR1}(a) are followed in Fig.\ref{fig:EPR_fig2}(a,b) at $10$\,K. The calculated magnetic moment for Tm2 within our CF model is also shown in Fig.\ref{fig:EPR_fig2}(a), the result was shifted to schematize the effect of the symmetry operations linking the four magnetic sites (see Fig. \ref{fig:Struct}(a)). Fig.\ref{fig:EPR_fig2}(b) shows the g-factors extracted from the red and black modes in Fig.\ref{fig:EPR1}(c) by rotating in the $(HHL)$-plane.  These modes are well described by the space-group symmetry and shows that the assumption of a planar anisotropy for Tm2 describes well the EPR signal.

Fig.\ref{fig:EPR_fig2}(c) shows the angular dependency of the extracted g-factors for Tm1 at $40$\,K and $10$\,K, which were assigned with blue and green arrows/lines in Fig.\ref{fig:EPR1}(a) and Fig.\ref{fig:EPR1}(c). It is important to notice that it is not possible to follow the angular dependency of these signals within the $ab$-plane, most likely due to very small g-factor values. The inset in Fig.\ref{fig:EPR_fig2}(c) shows the calculated magnetic moment for Tm1 within the ab-plane according to our CF model. The average (bulk) response would be almost constant when considering the symmetry operation linking the magnetic sites, appearing as a planar anisotropy made out of shifted 4 sites with Ising anisotropy.  The large extracted g-factor for Tm1 with B$_\parallel$c agrees well with Tm having $3+$ oxidation state and shows that the admixed pseudo-doublet has a total angular momentum quantum number bigger than $\mathbf{J}_z=6$ (for $\mathbf{J}_z=6$, g$_\parallel$=14).  

\section{Conclusions}

We have shown experimental evidence of the different and co-existent single-ion anisotropies in \stm. The ground state can be considered as a pure pseudo-doublet at least for Tm1 and an admixed pseudo-doublet for Tm2, as this site is not limited to an Ising anisotropy as described in the effective spin formalism developed in Refs. \cite{Rau, TmMgGaO4_theory}.

The magnetic excitations were measured at a temperature above the peak in the specific heat, therefore we described the Hamiltonian in terms of the total angular momentum quantum number. It is interesting to notice that although no interactions between the chains are needed in our model, the dispersion relation is the same for both sub-lattices due to the similarity in exchange path and $\alpha$ values. The energy gaps appear larger in the data than in our model, perhaps due to inaccuracies in the CF model or additional contributions to the gap size arising from frustration effects. 
The magnetic exchange interactions reported here are similar to those in \sho \, reported in Ref.\cite{srho2o4} where no interaction between the different magnetic sublattices is considered. However, in the thulium case both $\mathcal{J}_1-\mathcal{J}_2$ interactions are AFM.  

The ground-state phase diagram for the $\mathcal{J}_1-\mathcal{J}_2$ contains very rich physics for both the Ising and planar cases. Vector-chiral phases, Néel phases and dimer phases are calculated as ground states in the planar case using multiple calculational tools (see e.g. Ref.\cite{Sato}). The excitations in the fully frustrated case ($\mathcal{J}_1/\mathcal{J}_2\approx 1 $) are expected to be incommensurate, as is the case in \stm, while the ground states remain paramagnetic.  A gapped quantum paramagnet is also expected in the Ising case \cite{Jalal}. Although there are various physical realizations of frustrated  $\mathcal{J}_1-\mathcal{J}_2$ with AFM and FM interactions, such as the quasi-1D edge-shared cuprates \cite{Ueda, Grafe}, not many examples of AFM-AFM $\mathcal{J}_1-\mathcal{J}_2$ frustrated chains exist, being \stm \, an unusual case.  Further investigation of the
magnetic excitations in the $S = 1/2$ regime in the presence of a magnetic field is key to comparing calculated phase diagrams and critical parameters. 
 
\section{Acknowledgments}
We acknowledge fruitful discussions with M. D. Le and J. Rau.  This work is based on experiments performed at the Swiss spallation neutron source SINQ, Paul Scherrer Institute, Villigen, Switzerland, accessed through the Norwegian Center for Neutron Research – NcNeutron (The Research Council of Norway, project no. 245942). We acknowledge the support of HLD at HZDR, a member of the European Magnetic Field Laboratory (EMFL).
R.T.P would like to thank the Danish Agency for Science and Higher Education for its support under DANSCATT. Parts of this work were financed through the Norwegian Council (project number 324863).

\end{document}